\documentclass[sigconf]{acmart}

\usepackage{stfloats}

\newcommand{\modified}[1]{\textcolor{black}{#1}}
 
\usepackage{array}
\usepackage{booktabs}
\AtBeginDocument{%
  \providecommand\BibTeX{{%
    \normalfont B\kern-0.5em{\scshape i\kern-0.25em b}\kern-0.8em\TeX}}}

\setcopyright{acmcopyright}

\copyrightyear{2024}
\acmYear{2024}
\setcopyright{rightsretained}
\acmConference[DIS '24]{Designing Interactive Systems Conference}{July 1--5, 2024}{IT University of Copenhagen, Denmark}
\acmBooktitle{Designing Interactive Systems Conference (DIS '24), July 1--5, 2024, IT University of Copenhagen, Denmark}\acmDOI{10.1145/3643834.3661530}
\acmISBN{979-8-4007-0583-0/24/07}

\acmSubmissionID{2218}

\begin{document}

\title[Tangible Scenography as a Holistic Design Method for HRI]{Tangible Scenography as a Holistic Design Method for Human-Robot Interaction}

\author{Amy Koike}
  \authornote{The first two authors contributed equally to this research.}
\orcid{0009-0000-9088-6578}
\affiliation{
  \institution{University of Wisconsin--Madison}
  \city{Madison}
  \state{Wisconsin}
  \country{USA}
  \postcode{53706}
}
\email{ekoike@wisc.edu}

\author{Bengisu Cagiltay}
\authornotemark[1]
\orcid{0000-0001-7024-4957}
\affiliation{%
  \institution{University of Wisconsin--Madison}
  \city{Madison}
  \state{Wisconsin}
  \country{USA}
  \postcode{53706}
  }
\email{bengisu@cs.wisc.edu}

\author{Bilge Mutlu}
\orcid{0000-0002-9456-1495}
\affiliation{%
  \institution{University of Wisconsin--Madison}
  \city{Madison}
  \state{Wisconsin}
  \country{USA}
  \postcode{53706}
  }
\email{bilge@cs.wisc.edu}

\ccsdesc[500]{Human-centered computing~Scenario-based design}
\ccsdesc[500]{Human-centered computing~Participatory design}
\ccsdesc[500]{Human-centered computing~User centered design}
\ccsdesc[500]{Human-centered computing~Activity centered design}
\ccsdesc[500]{Human-centered computing~User studies}

\begin{abstract} 
Traditional approaches to human-robot interaction design typically examine robot behaviors in controlled environments and narrow tasks. These methods are impractical for designing robots that interact with diverse user groups in complex human environments. 
Drawing from the field of theater, we present the construct of \textit{scenes}---individual environments consisting of specific people, objects, spatial arrangements, and social norms---and \textit{tangible scenography}, as a holistic design approach for human-robot interactions. 
We created a design tool, \textit{Tangible Scenography Kit (TaSK)}, with physical props to aid in design brainstorming. We conducted design sessions with eight professional designers to generate exploratory designs. Designers used tangible scenography and TaSK components to create multiple scenes with specific interaction goals, characterize each scene's social environment, and design scene-specific robot behaviors. 
From these sessions, we found that this method can encourage designers to think beyond a robot's narrow capabilities and consider how they can facilitate complex social interactions.

\end{abstract}

\keywords{Tangible scenography, human-robot interaction, design research}

\begin{teaserfigure}
    \includegraphics[width=\textwidth]{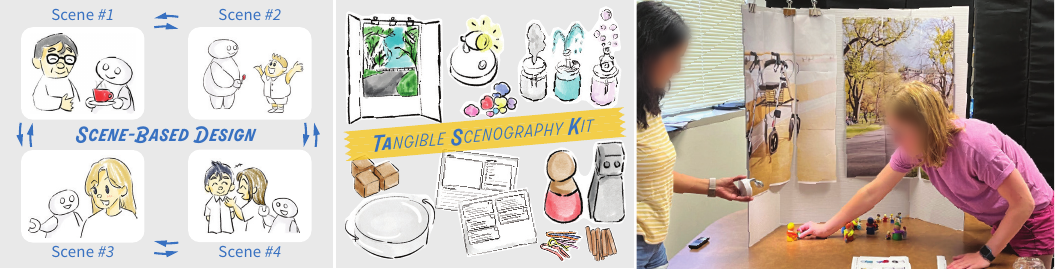}
  \caption{\textit{Tangible Scenography as a Holistic Design Method} --- In this work, we follow a scene-based design approach (left) for creating scenarios in HRI. To support scene-based design, we developed \textit{Tangible Scenography Kit (TaSK)} that includes physical props and effects inspired by theater production (center). To explore how our proposed method \textit{tangible scenography} can support the process of creating holistic human-robot interaction scenarios, we applied this method with designers (right).}
  \Description{A photo on the left. An illustration is on the right. The photo on the left shows a person who is holding a figurine and another person who is holding a spotlight. The illustration on the right includes four panels, and each panel includes a scene where a robot interacts with people.}
  \label{fig:teaser}
  \vspace{3pt}
\end{teaserfigure}

\maketitle

\section{Introduction}

    Social robots, as physically embodied agents, pose unique design challenges in the wild.
    As social robots navigate within complex real-world settings \cite{lee2012personalization}, they need to actively engage with users by emulating aspects of human social interaction that include speech~\cite{lucas2017role}, gaze~\cite{admoni2017social}, proxemics~\cite{mead2017autonomous, takayama2009influences}, gestures~\cite{breazeal2005effects}, and facial expressions~\cite{saldien2010expressing}.
    Current design practices tailored for interactions between humans and robots, as noted by~\citet{lupetti2021designerly}, include generative methods to conceptualize robot behaviors and appearances~\cite{walters2007exploring, li2023want} that draw inspiration from interdisciplinary fields including animation~\cite{schulz2019animation, desai2019geppetto, sirkin2015mechanical, vazquez2014spatial}; sketching and 3D modeling~\cite{hoffman2014designing, tonkin2018design, alves2017yolo, sakamoto2009sketch}; and storyboarding, brainstorming, and bodystorming~\cite{rose2017designing, bjorling2019participatory, porfirio2019bodystorming, abtahi2021presenting}; as well as methods for research through design, such as speed dating or user enactments~\cite{luria2019championing}, end-user programming~\cite{senft2022participatory}, tangible design~\cite{porfirio2021figaro, guo2009touch}, and rapid prototyping~\cite{huang2016design}. 
    
    Prior approaches to designing human-robot interactions, including the methods discussed above, include powerful approaches that capture the nuances of social interaction and iteratively refine robot designs based on user feedback, cultural context, and specific application domains. Despite the ability of these approaches to create sophisticated, multimodal robot behaviors, they have primarily focused on singular interactions within confined environments. This focus, which might be suitable for technologies with narrow or limited use cases, may fall short in robots designed to work in complex human environments. 
    
        Consider designing a care robot placed in an assisted living facility to make deliveries of mail, medication, food, beverages, and supplies to resident rooms. A task-based approach might focus on how the robot will navigate to a given room, announce its arrival, and make its delivery. However, this approach is limited in considering the real-world interactions that may emerge spontaneously, such as casually offering help to visitors searching for a room, playfully engaging with children curiously approaching the robot, or stopping by the lobby to join others welcoming a new resident to the facility. Hence, leading to missed design opportunities that could explore how the robot might interact in these complex settings. Designing for these rich interactions, and even envisioning the kinds of spontaneous engagements a robot navigating in a complex human environment, requires new design approaches and methods that consider a more holistic representation of the design space for real-world human-robot interactions.
        
    In this paper, we introduce \textit{scene-based design} as a holistic design framework inspired by the field of theater and scenography. Scene-based design considers human-robot interaction scenarios as a set of scenes that can emerge spontaneously, similar to how theater and scenography are composed of multiple scenes to express a story.
    \modified{We propose the concept of \textit{``tangible scenography''} as a method for scene-based design. To support this approach, we developed a design artifact, \textit{Tangible Scenography Kit (TaSK)}. We report insights from applying this method with eight professional designers.}
    \modified{
    Our work is motivated by the following design challenges:
    \begin{itemize}
        \item \textbf{Design Challenge\#1:} What \textit{design approaches} can support creating holistic HRI scenarios that represent dynamic human environments and a wide range of robot interactions?
        \item \textbf{Design Challenge\#2:} What \textit{design resources} may be useful in supporting the practices of creating holistic HRI scenarios?
        \item \textbf{Design Challenge\#3:} How can these methods and resources \textit{facilitate designers' process} of creating holistic HRI scenarios?
    \end{itemize}
    }

In response, we make the following contributions:
\begin{itemize}
    \item \textbf{Design Method:} We propose \textit{tangible scenography} as a holistic design approach for \textit{scene-based design,} inspired by the fields of theater and scenography.
    \item \textbf{Design Artifact:} We developed a scene-based design artifact, called \textit{Tangible Scenography Kit (TaSK)}, containing accessible, tangible, and modifiable components including stage effects, scenic design, and craft supplies.
    \item \textbf{Empirical Understanding:} We conducted exploratory design sessions with eight professional designers. Participants provided insight into how tangible scenography and TaSK supported their \textit{process} of creating holistic HRI scenarios.
\end{itemize}

\section{Background}

\begin{figure*}
    \centering
    \includegraphics[width=\textwidth]{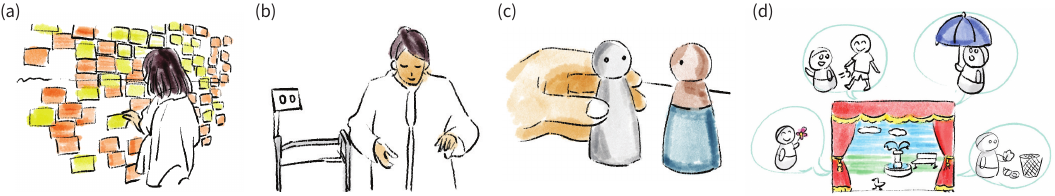}
    \caption{\textit{Methodological Background for Tangible Scenography:} Our proposed method contribution (see (d))-- is contextualized within the broader landscape of design methods in HCI, including \textit{design thinking} and \textit{design exploration} (see (a) adapted from \cite{luria2020robotic}); \textit{performing arts} (see (b) adapted from \cite{luria2020robotic}); and \textit{tangible design toolkits and props} (see (c) adapted from \cite{porfirio2021figaro})}
    \Description{Four illustrations are aligned. The first image from the left shows a person who is brainstorming by using Post-it on a whiteboard. The second image from the left shows a person and a robot, which are acting on a stage. The third image from the left shows two figurines on a table. The last image shows the concept of scene-based design. The image shows a theater stage in the center and four bubbles. Each bubble has a scene that includes a robot: the robot is giving a flower to someone, the robot is playing with children, the robot is holding an umbrella, and the robot is picking garbage.
}
    \label{fig:RW-Motivation} 
\end{figure*}

Tangible Scenography, a scene-based design method introduced in this paper, draws inspiration from performing arts and transfers relevant knowledge to the field of human-robot interaction (HRI). To contextualize scene-based design within the broader landscape of design methods, we cover literature that spans across HCI and design thinking, performing arts, and toolkits for design exploration (see Figure \ref{fig:RW-Motivation} for a summary).

\subsection{Design Thinking Methods}

The model of research through design (RtD) in HCI~\cite{zimmerman2007research} captures the interplay between HCI researchers and practitioners. The process of ideation, iteration, and critique between the design researchers (\textit{true} knowledge) and engineers (\textit{how} knowledge) can lead to the production of the \textit{``right''} artifact to be delivered to HCI practitioners. Similarly, RtD and exploratory design practices in HRI can enable the design of \textit{``the right thing''}~\cite{luria2019championing}. \modified{More recently, there has been increased advocacy toward incorporating user perspectives and needs as part of robot interaction design methods~\cite{kim2011user,rogers2022maximizing, bjorling2019participatory}. However, the subjective and unstructured nature of RtD proposes challenges for its application in HRI~\cite{luria2019championing}. 
Hence, there is a gap in available HRI design methods that embrace the subjective and unstructured nature of RtD while still allowing designers, domain experts, and engineers to make practical design choices. 
With tangible scenography, we propose a design method that can address this gap and serve as the right level of abstraction to provide practical insights as an application of RtD.}

Furthermore, practice-based views on design~\cite{goodman2011understanding}, contextualized as ``designerly ways of knowing,'' are increasingly becoming common in the HRI field~\cite{lupetti2021designerly}. \modified{Common interaction design methods in HRI include scenario-based design~\cite{malik2015human,xu2015methodological, zlotowski2011interaction}, sketching and story-boarding~\cite{rose2017designing, tonkin2018design, sirkin2016design}, bodystorming~\cite{porfirio2019bodystorming, abtahi2021presenting}, 3D-modeling~\cite{hoffman2014designing, alves2017yolo, sakamoto2009sketch}, and prototyping~\cite{zamfirescu2021fake, krupke2018prototyping}. Less common but effective methods for exploring more naturalistic scenarios in HRI include narrative \cite{koay2020narrative, syrdal2014views} and playful approaches~\cite{lee2020ludic} to interaction design, as well as realistic in-situ prototyping~\cite{vsabanovic2014designing} and scenarios created from human-human interactions~\cite{zlotowski2011interaction, sauppe2014design}. For example, these playful approaches combine methods such as role-playing games~\cite{collins2021does}, card-games~\cite{schwaninger2021you}, or multi-player games~\cite{munoz2021robo}. Applications of these practice-based design methods in HRI typically narrow down the focus to capture a singular use-case scenario, with a pre-determined user persona, and a set of sequential events. While this approach may be practical in determining specific robot features in an isolated scenario, real-world interactions between people and robots will likely represent more complex and dynamically changing factors that cannot be captured fully within a scenario-based method. 
Hence, there is a gap in HRI design methods that are structured enough to capture practical HRI scenarios, as well as flexible enough to explore spontaneous, unstructured ecological changes that might happen in a real-world interaction between people and robots.}
Tangible Scenography as a holistic and exploratory design method in HRI is situated within this complex space. Overall, we seek to extend the practices of design thinking in HRI. To do this, we offer a semi-structured design approach that supports the ideation process of exploring scenarios that capture complex interactions between robots, humans, and their surroundings.

\subsection{Performing Arts and Interaction Design} \label{sec:background:performingarts}
Rich traditions of performing arts have informed the creation of interactive systems in HCI that resonate with human sensibilities (\textit{e.g.,} ~\cite{iacucci2002everyday, newell2006use, brandt2000evoking}). HRI researchers often transfer knowledge from creative arts and artists~\cite{workshop_arts, lehmann2013artists}, animation~\cite{schulz2019animation, desai2019geppetto, sirkin2015mechanical, vazquez2014spatial}, dance~\cite{cuan2018curtain, abe2022beyond}, choreography~\cite{can2016robot}, and improvisation~\cite{troughton2022robotic}. For example, inspired by shadow puppetry, \citet{porfirio2021figaro} designed a tabletop system for end-users to easily and tangibly author human-robot interaction programs. \citet{ryohei2013evaluation} designed an interactive puppet theater to entertain deaf children and invite them to participate in the show. 
\modified{Other examples of performing arts in HRI include co-performing with trained actors and robots. For example, a theater actor collaboratively performs with robots on a script that focuses on the relationship between elderly patients and their robotic caregivers~\cite{lu2017sky, lu2011human}. 
Similarly, \citet{chatley2010theatre} proposes ``Theatre HRI'' as a method that includes theatrical scenes being acted by professional actors interacting with robots, displayed in front of an audience. }
To scaffold design dialogues in a similar context of designing for at-home care, \citet{vines2014experience} proposes ``Experience Design Theatre'' as an HCI method, which brings together caregivers, theater experts, engineers, and designers to co-design and co-produce a non-scripted live theatre show.
In collaboration with designers and theatre experts, \citet{luria2020robotic} co-designed an immersive performance with robots to explore theatre performance as a form of knowing through doing. 
\citet{gemeinboeck2021aesthetics} introduces designing with ``bodying-thinging'' as a relational performative design approach that enables early-stage design for human-robot encounters. This approach embraces abstract robotic artifacts, such as a human-sized robotic cube, and transforms them into an active participant in social interactions.
\modified{Our proposed method, \textit{tangible scenography}, expands the focus beyond the performative actions and includes the scenic components that construct the performance. In other words, while current theater-inspired methods in HRI draw insights from performing arts and acting, our proposed method goes beyond the focus of acting but draws metaphors from the field of scene design, \textit{i.e.,} scenography. Tangible scenography borrows several elements from this artistic discipline, including props, backdrops, lighting, and sound effects, as well as methods such as script analysis, visual research, or prototyping. We incorporate these insights and principles from scenography to offer a framework that enables designers to explore and brainstorm more holistic, complex, and dynamic representations of human-robot interaction scenarios.}

\subsection{Toolkits and Props Used in Design Exploration}
Just as theatrical productions use props to convey emotions, narratives, and experiences to an audience, the tools and props employed in interaction design can help shape, guide, and enhance the creative process. A practical example of this includes using LEGO bricks as an alternative to paper-based prototyping ~\cite{de2022lego, bourdeau2020design, cagiltay2020investigating}. Researchers have explored the use of such props and toolkits as part of design exploration. For example, to support co-design sessions with children, \citet{walsh2013facit} proposes ``bags of stuff'' which includes low-tech prototyping supplies such as felt, glue, or feathers. \modified{Similarly, \citet{lee2022unboxing} applied the bags of stuff method when co-designing unboxing experiences for social robots with children.} \citet{forsslund2015woodenhaptics} created an open-source starter kit to craft haptic devices that are low-cost and easy to fabricate. \citet{mellis2013microcontrollers} used simple resources such as paper, conductive ink, and electronic components to craft circuits to support the learning of embedded programming. 
%
%
Moreover, \citet{alves2022flexi} created a robot embodiment kit, Flexi, that is a cost-effective customizable design kit with materials and attachments that allows its users to create appropriate designs for social robots easily. This kit is customizable enough to create social robot embodiments that may fit in various contexts, for example, as community support, mental health support, or education.
We are inspired by these several toolkits and props used in design exploration. Hence, our goal with the proposed \textit{Tangible Scenography Kit (TaSK)} is to similarly provide a set of accessible, cost-effective, and customizable resources that can support creative exploration in human-robot interaction design.

\section{Development of the Method}\label{sec:designprocess}

\modified{
In this section, we explore the first and second design challenges; \textit{\textbf{DC\#1:} What design approaches can support creating holistic HRI scenarios that represent dynamic human environments and wide range of robot interactions?} and \textit{\textbf{DC\#2:} What \textit{design resources} may be useful in supporting the practices of creating holistic HRI scenarios?}
}

\subsection{Design Inspiration} 
    To create a holistic design experience, we draw inspiration from the field of theater. This field typically creates imagined spaces --a set of scenes-- to express a story. 
    Similarly, we define a scenario, which is often referred to as an outcome of HRI design, as a collection of scenes. This interpretation of a scenario can create a holistic design experience that enables a designer to imagine the same robot in different scenes behaving differently. We call this concept \textit{``scene-based design.''}
    %
    Nevertheless, the challenge remains in creating a design environment to support designers in crafting holistic human-robot interaction scenarios that capture a variety of scenes. To address this, we draw inspiration from theater production and scenography. Theatrical productions employ props and stage effects (such as lighting, costume, sound, and projection design) to construct immersive and context-rich settings. 
    Acknowledging the physical --and tangible-- aspects of theatre enables the audience to immerse themselves in the narratives of each scene. We posit that these elements can likewise enrich the process of scene-based design, proposed as \textit{tangible scenography}.
    To enable \textit{tangible scenography}, we curated a resource kit including several props and mimicry items, aiming to provide designers with versatile tools for scene-based design.
    Our vision materializes in the form of the \textit{Tangible Scenography Kit (TaSK)} — a toolset specifically designed to support scene-based design. With this kit, we aim to establish an immersive environment, enabling designers to immerse themselves in human-robot interaction scenes, facilitate their creative processes, and communicate their ideas effectively. 

\subsection{Domain Expert Interview} 
     To better understand the typical process of scenography in the performing arts field we interviewed a domain expert in theatre production and scenic design. 
     The domain expert was a professor of scenic design at a university in the United States. We refer to this expert as SD in the following sections. SD has a degree in master of fine arts in scene design and an undergraduate degree in theater. SD worked for 2-years as an entertainment and experience designer, 5-years as a scenic designer working over 100 productions, and 2-years as an art director for television shows. From this interview, we identified key components for scene-based design and TaSK. 

    \subsubsection{Procedure} 
    To supplement our interview with SD, we presented materials including a slide show summarizing the motivation behind our method, a 3D-printed robot as a representation, and supplemental resources to demonstrate the expressiveness of the robot. 
    Our interview with SD lasted for one hour. Our goal was to understand their process of scene design within theater production and discuss how methods from theater and scenography could transfer to human-robot interaction design. Thus, we asked semi-structured interview questions in the following categories: (1) design process, (2) collaboration with other designers, (3) technology used, and (4) recommendations for transferring knowledge from theater production to human-robot interaction design. 
    Some example questions were: \textit{What is your typical process for developing scenes for a show, whom do you collaborate with, what resources and technologies do you use, how do stage design and props influence scenario creation of a play, what type of technology inspires your scene design process, which design elements from scene design may transfer to robot interaction design?} 

    \subsubsection{Analysis and Findings} 
    The first two authors conducted the interviews. Audio data from the interview session was collected and transcribed with automated transcription software and manually revised for accuracy. The first two authors collaboratively and iteratively coded the transcriptions and used affinity mapping as a method to identify key insights for relevant scenography practices.

    We found that, for SD, scene design begins as a conceptual process and gradually becomes more specific. SD initiates their design with \textit{script analysis}, expanding their imagination based on textual information. Following the script analysis, they transition to \textit{visual research}, considering how to visually convey the script's essence. Once scene designers have a clear vision of their intended visual expression, they start \textit{prototyping}, which serves both as an ideation phase and a means of communication with other collaborators, such as a director or a technical supervisor. Throughout these steps, scene designers also consider \textit{blocking}, where they specify actors' movements on the stage for the performance. 
    These insights into the design process informed our decisions for developing ``tangible scenography'' as a scene-based design method in HRI. We transfer this knowledge to our method with the following guidelines:
    \begin{itemize}
        \item \textbf{Scalable Design}: Scene-based design should be scalable enough to afford transitions between conceptual (high-level) and specific (low-level) design processes.
        \item \textbf{Textual and Visual Representations}: Scene-based design should facilitate text and visual-based processes, such as script analysis, storyboarding, and visual research methods.
        \item \textbf{Prototyping}: Scene-based design should support ideation by providing rapid prototyping tools.
        \item \textbf{Interdisciplinary Collaboration}: Scene-based design should enhance collaboration between designers and stakeholders (\textit{e.g.,} engineers, researchers, end-users).
    \end{itemize}

    Additionally, SD shared the following suggestions for a \textit{design kit}: \textbf{provide a shell for the story; have a prompt or a theme for the design kit; set the motivation for the interaction; balance between props and set decoration; provide basic elements of a composition as well as their counterparts} (\textit{e.g.,} for lighting, light and dark options.)

\begin{figure*}[!t]
    \includegraphics[width=\linewidth]{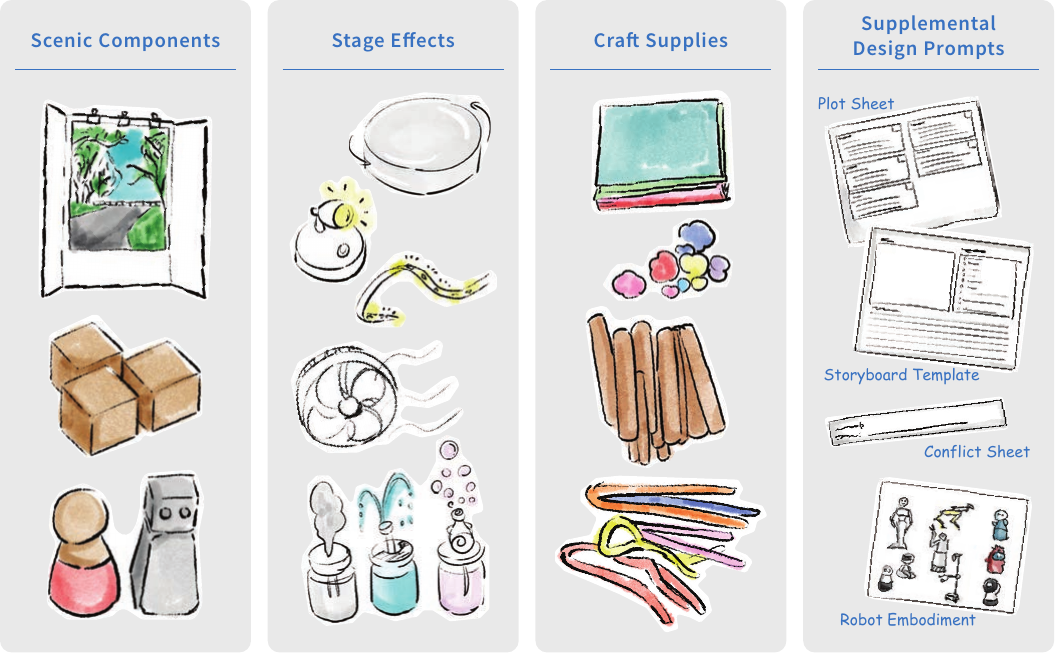}
    \caption{ \textit{Tangible design resources provided in our design artifact, Tangible Scenography Kit} The kit included \textit{scenic components} (backdrops, figurines, and blocks); \textit{stage effects} (fluid generation machines, spotlights, LED strips, turntable, and sound effects), \textit{craft supplies} (papers, felt, wooden sticks, pipe cleaners, and stickers), as well as \textit{supplemental design prompts} (a design brief including five contextual plot options, a conflict for the plot, a sheet illustrating robot embodiment, and a storyboard template).
    }
    \Description{Four panels show four categories of TaSK. The first group (top left) shows scenic components including a backdrop, wooden blocks, and figurines. The second group (top right) shows stage effects including LED strips, a fan, fluid expression machines, spotlights, and turntable. The third group (bottom left) illustrates craft supplies including paper, popsicle sticks, pipe cleaners, and stickers. The fourth group (bottom right) illustrates a set of papers.}
    \label{fig:materials}
    
\end{figure*}

\subsection{Design Process of TaSK} 

    TaSK (see Figure \ref{fig:materials}) has four categories: s\textit{cenic components, stage effects, craft supplies, and a supplemental design brief} (Table \ref{table:plot}). 

    Through a process of rapid and iterative prototyping, we identified that the kit should primarily consist of two key elements: scenic components and stage effects. \textit{Scenic components} set the context of the story by indicating where it takes place, often supported by backdrops. Props and costumes can also serve as scenic components, as they contribute to additional contextual details. In contrast, \textit{stage effects} introduce more abstract elements to establish an atmosphere or mood through the use of color, sound, or special effects such as fog. For instance, purple-colored lighting and a fog-filled stage can create a spooky impression among the audience. Thus, we decided to include both types of elements and brainstormed potential components for the kit.
    
    We ultimately selected: robot-like figurines, human-like figurines, customizable LEGO figurines, building blocks, and contextual backdrops as \textit{scenic components}; and fluid effect devices (\textit{i.e.,} to generate fog, wind, bubbles, and water droplets), light, sound effects, and a turntable as \textit{stage effect components}. The reasons for selecting these specific stage effect components include their similarity to broadly used components across theatrical shows and the ability to easily be purchased or built. These two reasons were important because we envision our kit to be available and accessible as an open-source and tabletop-scale resource for HRI researchers and designers. The scenic components were selected with the same reasoning, however, thematic props and costumes were removed because they might be too specific to a narrow context and may constrain designers' creativity or imagination. As an alternative, we decided to include \textit{craft supplies} as a part of our kit to enable the creative freedom of designers. 

    The supplemental design prompts, \textit{i.e., the design brief} (partly in Table \ref{table:plot}), included a plot sheet, a conflict sheet, a storyboard template, and a sheet illustrating varying robot embodiment.
    %
    When developing the \textit{plots}, we included common applications where robots are currently used or envisioned for future use.     
    Each plot included a role for the robot (\textit{i.e.}, patrol, guide, companion, education, and delivery), a context, two locations where the robot could be present, and a motivation of the robot in the context. 
    The \textit{conflict} sheets introduce a conflict arising from either \textit{user interactions} or \textit{environmental factors} tied to each plot. For example, a conflict might involve a child \textit{(user)} kicking the robot while it is interacting with children, or the weather conditions \textit{(environment)} changing while the robot is on patrol. 
    To support sketching and note-taking, we designed a \textit{storyboard template} for participants. The template included a checklist of kit components, a blank panel for sketching, and lines for note-taking. The template was also used by researchers to track the design process and take observation notes. 
    Additionally, we provided a sheet with \textit{robot embodiments}, which displays pictures of robots categorized into three types: mobile \& social robots, mobile robots with grippers, and stationary social robots.
    The conflict and robot embodiment sheets were designed to provide semi-structured milestones within the design session, as well as fostering ``what-if'' discussions to promote exploratory ideation. 

\begin{table*}[!t]
    \caption{\textit{The design brief:} Includes options for the robot's role, context, location, motivation, and conflict sheet.}
    \Description[Table]{The design brief is shown in the table with six rows and four columns.} 
    \label{table:plot}
    \centering
    \small
    \begin{tabular}{lp{0.22\linewidth} p{0.15\linewidth} p{0.18\linewidth} p{0.27\linewidth}}
        \toprule
    \textbf{\textit{Role}} & \textbf{Context} & \textbf{Location} & \textbf{Motivation} & \textbf{Conflict (introduced separately)} \\
    \toprule
    \textit{Patrol} & The robot is patrolling an area. & Public Park OR Outdoor Mall & Identify littering on the streets. & There is a sudden change in the weather climate. \\
    \hline
    \textit{Guide} & The robot is guiding people. & Eldercare Facility OR Center for Visually Impaired & Small talk while guiding people between rooms. & The user feels tired and wants to sit down. \\
    \hline
    \textit{Companion} & The robot is playing with children & Bedroom OR Living room & Encourage children to tidy up the room. & Child kicks the robot. \\
    \hline
    \textit{Education} & The robot is helping a teacher with a science experiment. & Class OR Museum & Increase classroom engagement. & A participant raises their hand for a question but the teacher does not see them. \\
    \hline
    \textit {Delivery} & The robot is delivering a package & Office OR Hospital & Deliver a package from pickup to drop off location. & Something prevents the robot from moving forward. \\
    \bottomrule
    \Description{The list of design briefs is shown in the table with six rows and five columns.}
    \end{tabular}
\end{table*}

    The key components of the finalized kit are provided for open access\footnote{Details of the expert interview materials, kit component materials, and procedures for the design sessions, can be found at the following \textit{Open Science Foundation (OSF)} repository: \url{https://osf.io/au9x8}} and are summarized as:
    \begin{itemize}
        \item \textbf{Stage effect components}: Fluid generation machines, spotlights, LED strips, turntable, and sound effects;
        \item \textbf{Scenic components}: Backdrops, figurines, and blocks;
        \item \textbf{Craft supplies}: Papers, felt, wooden sticks, pipe cleaners, and stickers;
        \item \textbf{Supplemental design brief}: A design brief including written prompts capturing a plot, conflict, figures of robot embodiment, and a storyboarding template.
    \end{itemize}

    \begin{table}[!b]
    \caption{Participant backgrounds and demographics}
    \Description[Table]{The participants' background information is shown in the table with nine rows and four columns.}
    \centering
    \small
    \label{table:background}
    \begin{tabular}{p{.5cm}p{5cm}p{.5cm}p{1cm}}
    \toprule
    \textbf{ID} & \textbf{Background} & \textbf{Age} & \textbf{Gender} \\ \toprule
    P1 & Artist; paper making, sculpture, educator& 65& Female   \\ \midrule
    P2 & College educator; computer science, dramatic art and performance studies.  & 62  & Male \\ \midrule
    P3 & Landscape architecture design & 28                       & Female \\ \midrule
    P4 & Graphic design& 22& Agender \\ \midrule
    P5 & Book editor; museum exhibits. studio arts; printmaking, photography, graphics design    &  60 & Female \\ \midrule
    P6 & Video game artist. Studio arts; 2D/3D art creation  & 25  & Female \\ \midrule
    P7 & Sr user experience designer   & 40 & Female \\ \midrule
    P8 & Communications specialist; visual communication production  & 26 & Male \\ \bottomrule
    \end{tabular}
    \Description{The participants' background information is shown in the table with nine rows and four columns.}
\end{table}

\section{Exploration of Tangible Scenography as a Design Method}\label{sec:designsession}
\modified{  
    We conducted exploratory design sessions to address the third design challenge, \textit{\textbf{DC3:} How can these methods and resources facilitate the design process of creating holistic HRI scenarios?} In these design sessions we observed how designers approach creating human-robot interaction scenarios applying \textit{tangible scenography}. 
    }

    \subsection{Participants} Eight designers (2 male, 5 female, 1 agender), aged 22--65 ($M = 41$, $SD = 17.2$), participated in a single session that lasted 90 minutes. Participants were recruited through mailing lists and were screened for moderate to high experience in design background. \modified{Participants did not have any professional experience with robots and the human-robot interaction field. }Table~\ref{table:background} summarizes participant backgrounds and demographics. We refer to these individuals as \textit{P1}--\textit{P8}. Participants received \$23 USD for each session.

    \subsection{Materials}
    The materials used in the study included the \textit{Tangible Scenography Kit (TaSK)}, with the stage effect components, scenic components, craft supplies, and design brief described in Section \ref{sec:designprocess}. Figure \ref{fig:setup} (a-b) illustrates the room layout and available materials in the study.

    \subsection{Procedure}
    Facilitators introduced the study goals and activities to the participants and obtained informed consent. Later, the participants were given the design brief (represented in Table \ref{table:plot}) and selected one contextual plot out of five options to design for.
    After selecting the contextual plot, participants picked the location where they wished the plot to take place, as well as a scenic background of their preference. Ten available scenic backgrounds were presented to the participants. Facilitators attached an enlarged printed version of the selected scenic background on the backdrop. 
    
\begin{figure*}[!t]
    \centering
    \includegraphics[width=\linewidth]{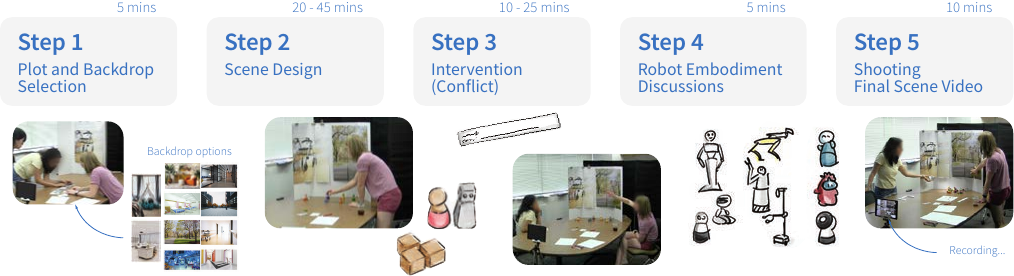}
    \caption{The \textit{tangible scenography} procedure consisted of five steps: (1) Plot and Backdrop Selection (5 mins) (2) Scene design (20-45 mins) (3) Planned Intervention (10-25 mins) to introduce a conflict in the plot (4) Discussion of Robot Embodiment Options (5 mins) (5) Final video recording to communicate the scene (10 mins). }
    \Description{An illustration shows the procedure for the scene-based design sessions. On the top, five gray colored boxes say step 1 to step 5. Under the boxes, there are four photos and several illustrations. The photo under the Step 1 box shows two researchers working with a participant. The photo under the step 2 box captures the participant holding a figure. The photo under step 3 captures a researcher and a participant discussing. The photo under step 5 shows the scene where a researcher and a participant are taking a video.}
    \label{fig:process}
\end{figure*}

\begin{figure*}[!t]
    \includegraphics[width=\linewidth]{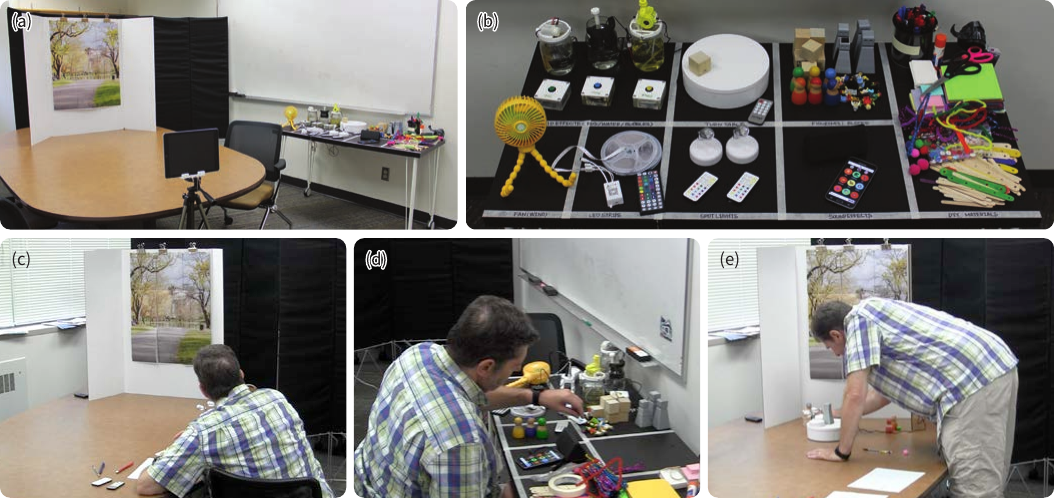}
    \caption{\textit{Materials and procedure for a tangible scenography.} (a) study room set up (b) the \textit{Tangible Scenography Kit (TaSK)} (c) Participant sketching, (d) exploring the kit components, (e) using the kit components to build a scene.
    }
    \Description{Two illustrations are aligned on the top and labeled (a) and (b) from the left, and their other photos labeled (c) to (e) are placed on the bottom. (a) shows a room layout for the design session, it captures a wide table, a backdrop on the table, a tablet standing on the side, and a table near the wide table. (b) shows what is on the smaller table. It has fluid expression devices and a fan on the left, LED, spotlights, a turntable, figurines, a smartphone in the middle, and craft supplies on the right. (c) to (e) capture a participant during the design session. (c) captures a participant sketching their idea on the paper, (d) captures a participant playing with the design kit components, and (e) shows a participant setting out the components on the wide table.
}
    \label{fig:setup}
\end{figure*}

    Participants were then prompted to \textit{create the scene} by applying their preferred \textit{brainstorming} method and using as many or as few of components from the design kit (\textit{e.g.,} \textit{``feel free to use either paper sketching or the toolkit to do some brainstorming for your scene.''})
    Participants described, discussed and iterated over the details of their scene with the facilitator, which typically ranged from 20-45 minutes. The facilitator then introduced a pre-prepared prompt introducing a \textit{conflict}. Given the new conflict, participants elaborated on their designed scenes, which typically ranged from 10-25 minutes. 
    Afterward, facilitators introduced a sheet illustrating \textit{robots} with different embodiment and capabilities, prompting the participants to discuss their scenes and reconsider the various robot capabilities. Finally, the participants recorded and narrated a short \textit{video} describing their final design, \textit{``to communicate to an engineering team.''} Once the video was recorded, participants were briefly interviewed about their experience during the study, how they connected their experience to the design session, and whether they had any feedback to improve this design method. Figure \ref{fig:process} and Figure \ref{fig:setup} (c-d-e) exemplifies the procedure.

\subsection{Data Collection and Analysis}
    Audio and video data were collected from the design sessions. 
\modified{
    We additionally recorded observation notes for each participant throughout the study. These notes documented various aspects of participants' design processes, including their interactions with the TaSK components, time allocation during design sessions, and any notable behaviors observed by experimenters. 
    During the interview sessions, these observation notes served as valuable reference points for clarifying ambiguous behaviors, allowing us to ask deeper participants' decision-making processes (\textit{e.g.,} probing questions such as \textit{``why did you choose that?''}) Furthermore, the observation notes provided supplementary material for our data analysis, offering nuanced insights into participants' thought processes and actions.
    }

    The first two authors were familiarized with the data by facilitating the design sessions, transcribing, and coding the transcriptions.
    For the analysis, authors conducted affinity diagramming on a digital whiteboard application and created transition diagrams for thematic analysis~\cite{braun2006using}. The transition diagrams were created to synthesize and visualize the participants' processes and approaches throughout the sessions. This was achieved by analyzing the video data annotated with the relevant quotes and statements from participants from the session transcripts. Triangulating these transition diagrams, video recordings, and audio transcripts, the authors identified four main themes that highlight participants' approaches enabled by scene-based design. The authors iteratively discussed and refined the identified themes reported in this paper.

\section{Insights from Tangible Scenography as a Scene-Based Design Method}\label{sec:results}

    We report four key insights reflecting our findings from the tangible scenography sessions (described in Section \S \ref{sec:designsession}): (1) \textit{insights from the participants' design process and output} --what designs emerged from tangible scenography, (2) \textit{participants' diverse design strategies}-- what types of strategies and approaches tangible scenography enabled, (3) participants' perceptions toward TaSK components, and (4) reflection on the \textit{role of facilitation} in the design process.

\subsection{Insight \#1: The Design Process}\label{sec:results:process}

\begin{figure*}[!t]
    \centering
    \includegraphics[width=\linewidth]{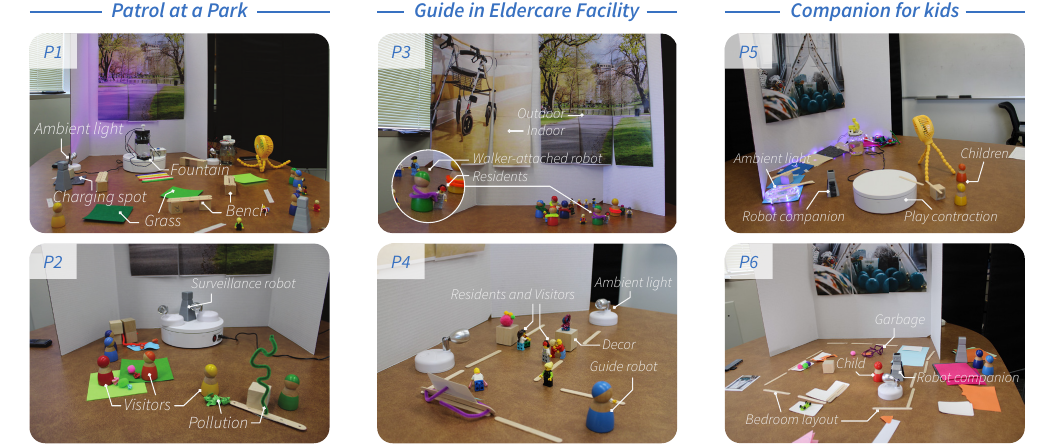}
    \caption{\textit{Participants' final scenarios from scene-based design sessions}: [P1] created a scenario for a \textit{``friendly neighborhood robot''}, including a rotating fountain, grass areas, benches, lighting, and a charging station in the park. [P2] created a \textit{``story of a teaching robot,''} depicting a surveillance robot with spotlights for non-verbal communication and pollution from park visitors. [P3] created a scenario to guide elderly residents from \textit{indoors to outdoors} and meet children at the park for therapy. [P4] created an indoor navigation scenario in an elder care facility, where the robot guides residents and conducts \textit{small-talk about decor and bulletin board}. [P5] and [P6] used play to motivate children to clean their bedrooms. [P5] created a social play scenario with positive reinforcement and [P6] created \textit{competitive and collaborative games}.}
    \Description{Six photos show the final state of participants’ scenarios from scene-based design sessions. All photos include a backdrop and the kit components on the table. The first photo (top left) includes a water droplet machine and a turntable in the center. There’s a spotlight which is purple color. Also, there is colored felt illustrating the grass area, and crafts made with wooden cubes and popsicle sticks to illustrate benches. The smoke machine and fan are placed on the edge of the table.
    The second photo (bottom left) has a turntable in the center. On the turntable, two spotlights and a robot figurine are placed. Along the turntable, there are crafts made with blocks and paper, and also figurines are placed. 
    In the third photo (top center), there are figures representing robots, elderly people, and preschool children.
    In the fourth photo (bottom center),  popsicle sticks are placed to make a hallway. In the hallway, LEGO figurines, art decors, and a bulletin board made with paper and pipe cleaner are placed.
    In the fifth photo (top right), felts are aligned near the backdrop. Several figurines, a turntable, and a fan are placed in the front of the backdrop.
    In the sixth photo (bottom right), popsicle sticks are placed to make a rectangle. In the rectangle, papers, figures, and a spotlight are placed. }
    \label{fig:final-scenarios}
\end{figure*}

    Out of the five available contextual plots, four of them were chosen by the participants: patrolling (P1 and P2); guiding (P3 and P4); children's companion (P5 and P6); and educational assistant (P7 and P8). Figure \ref{fig:final-scenarios} illustrates the final state of scenarios from each participant and Figure \ref{fig:scenes-P4} is an example of scenes from P4.
    
    We identified two milestones that were key to guiding participants' design process and their final output: \textit{opening scenes} and \textit{conflict scenes}. Most of the participants were able to produce complete scenarios that captured these milestones, except for two participants (P7 and P8). 
    \textit{Opening scenes} are the initial stages or starting points of the scenarios. These scenes typically represent the beginning of the robot's engagement with the users and surroundings. In their opening scenes, participants often described the setting of the stage, introduced characters, and defined the goals of these characters and stage components. 
    \textit{Conflict scenes} are the specific stages of the scenarios where a challenge, issue, or unexpected event arises that requires the robot to respond to or adapt its behavior. When responding to the conflict, we observed that participants either introduced new scenic components to demonstrate a change in the environment, or diverged and created parallel scenes capturing different behaviors or strategies for the robot. In this section, we describe each participant's process for creating their scenario and describe their final design.
    %

    \subsubsection{Patrolling Context} 
        P1 and P2 selected the patrolling scenario, and both chose the public park as the location. However, P1 and P2 demonstrated distinct narrative approaches in their design process. P1 constructed a story of a \textit{``friendly neighborhood robot''} that patrols the park and has friendly conversations with parkgoers. P1 crafted a rotating water fountain and park benches from the supplies and set the scene around them. The components set the ambiance and stage for the robot's interactions with park visitors, which included office workers, park visitors, a grandmother with grandchildren, and a skateboarder. P1 emphasized the robot's characteristics and engagements in social interactions, \textit{e.g., ``robot is going to probably have to have some kind of curiosity element to be able to have a conversation to begin with''} but also should \textit{``wander around and not bother anybody.''} P2's scenario, on the other hand, centered around \textit{``the story of the teaching robot,''} and used spotlights for the robot to deliver non-verbal communication and draw attention to environmental concerns. The robot's goals were to direct human awareness towards pollution and promote connection-making between strangers in the park. The scene included people engaged in soccer, cooking, and fishing while producing various pollutants. To illustrate the action of patrolling, P2 placed a robot figurine on top of the turntable and set it to traverse between 90-degree angles. 

        \textit{After the conflict was introduced (i.e., there is a sudden change in the weather)}, 
            P1 added three additional stage effects to the scene, the fog machine combined with the wind machine and spotlight color change, to illustrate a change in weather. P1 then continued their scene sequentially given the new conflict and simply alerted the parkgoers by physically approaching them. P2, however, created four different scenes for the robot's interaction: one focusing on the robot cleaning litter in the park after people have left; the second focusing on the robot actively and verbally promoting social change against littering; the third scene focusing on the robot's non-verbal behaviors to promote social change (via light direction and color change); and a fourth option focusing on a ``one-word'' message displayed on the robot's screen to communicate the problem of litter to people at the park to raise awareness and promote change.

    \subsubsection{Guiding Context} 
    P3 and P4 chose an eldercare facility for the guide scenario and approached it from different landscapes. P3 requested to design for both the facility's indoors and a nearby park. P3 combined these locations to allow the robot to guide elderly residents from indoors to outdoors to meet pre-schoolers to support rehabilitation. Drawing on their kinetic architecture expertise, P3 used craft supplies to create a dynamic walker bar around the robot for residents' mobility. P4 designed a corridor and background actors using TaSK supplies, including a wheelchair-bound woman conversing with a nurse, a walker-wielding man, a person exiting a room, and family members roaming the hallway. P4 expressed that their design was inspired by past elder-care visits: \textit{``It's a space that I kind of already know, pretty well. And I can kind of visualize potential, I guess, scenarios or things that could happen that the robot might have to deal with.''} P4 utilized TaSK to create decor options for the facility, such as decorating the hallway with popsicles sticks, wooden blocks, and craft materials, creating art fixtures, bulletin boards, and yellow mood lighting. P4 featured these components as conversation starters for elderly-robot interactions in their scenes.

    \textit{After the conflict was introduced (i.e., the user feels tired and wants to sit down)} 
        P4 branched out and created three parallel scenes focusing on different strategies for the robot: (1) the robot proactively takes the older adult to another room, following its internal map; (2) the robot asks the older adult how to continue, gives three options, and obeys the chosen option; and (3) the robot calls staff or asks for help from surrounding people. On the contrary, P3 simply continued their initial scene and further demonstrated how their designed ``dynamic walker'' would be used to resolve the conflict. P3 noted that this dynamic walker should be movable and strong enough to help the older adult stand up and sit down, and it should help guide the older adult without the need for a walking cane.

    \subsubsection{Children's Companion Context}
        P5 and P6 selected the child's companion scenario, and both participants similarly transformed the chore of tidying up a bedroom into a playful narrative. P5, influenced by their professional background of positive reinforcement principles and dog training practices, created two playful game options in which the robot employs motivational strategies for children to collaboratively and playfully clean their room. The first game depicted a collaborative mechanism where the child and robot could hand over the items to each other to place them in the respective storage areas, and the second depicted a solo game for the child to move the items to a storage area. In contrast, P6, a game designer, used various game mechanisms and motivational strategies to encourage tidying. P6 proposed six parallel scenes focusing on various game mechanisms (\textit{e.g.,} improv, collaborative, time-based, color-game, sound game, physical game). They then created two scenes that show how the robot can motivate compliant and non-compliant children. P6 then offered two more scene adaptations based on age or group formation.

        \textit{After the conflict was introduced (i.e., the child kicks the robot)}, 
            P5 created two scenes capturing different robot personas and strategies of motivation. Here, the robot may use either positive reinforcement (\textit{e.g.,} robot starts giggling and politely asks the child to help it to get up, encouraging helpful behavior) or aversive conditioning (\textit{e.g.,} repeating a loud noise or squirting water to the child, encouraging the child to clean up faster in a fun way). P6 offered three scenes capturing different robot strategies: (1) social-emotional learning, where the robot talks through its feelings and the child's emotions; (2) diversion, involving the robot telling the child that it hurts and motivating the child to continue cleaning up; and (3) reporting and asking for help from a parent if this behavior becomes repetitive.
        
    \subsubsection{Educational Assistant Context} 
        For the educational assistant scenario, P7 initially chose the museum and P8 chose the science class as their background. However, the scene-based design method - tangible scenography - presented distinct challenges for P7 and P8 in which both participants were limited in creating scenes. With the facilitator's intervention, both participants opted for alternative scenarios: the guide role in an eldercare facility (P7), and the delivery role in the office (P8). However, both participants still deviated from scene-based design and discussed tasks and functionalities for an in-home assistive robot for their elderly parent (P7) and an office-based trash-collecting robot (P8). Due to this divergence, these participants were not able to receive the conflict related to their plot. Their designs unfolded through verbal discussions and sketches and \textit{did not} include the design toolkit.

\begin{table*}
    \caption{\textit{Summary of design process and approaches of enabled by tangible scenography:} Participants selected a context to design for and applied different scene flow and brainstorming approaches to create an HRI scenario using various TaSK components.}
    \Description[Table]{A summary of the design process and approaches by our participants is shown in the table with nine rows and six columns.}
    \centering
    \small
    \label{table:components_used}
    \begin{tabular}{p{.6cm}p{1.5cm}p{1.5cm}p{2cm}p{3cm}p{6.75cm}}
    \toprule
    \textbf{ID} & \textbf{Context} & \textbf{Scene Flow} & \textbf{Brainstorming} & \textbf{Final Scenario} & \textbf{TaSK Components Used} \\ \toprule
    P1 & Patrol & Sequential & TaSK & Friendly
    & \textbf{Stage}: Water, Wind, Sound, Turntable, Spotlight \\ 
        &&&&Neighborhood& \textbf{Scenic}: Blocks, Figurines \\ 
        &&&&Robot& \textbf{Craft}: Fabric, Popsicle sticks \\ \midrule
    P2 & Patrol & Parallel &  TaSK  & The Story of the & \textbf{Stage}: Turntable, Spotlight \\
        &&& $\rightarrow$ Paper Sketching&Teaching Robot & \textbf{Scenic}: Figurine, Blocks \\
        &&&&& \textbf{Craft}: Pipe Cleaner, Furballs, Papers \\ \midrule
    P3 & Guide & Sequential & Paper Sketching & Indoor to Outdoor Guide
    & \textbf{Stage}: None \\ 
        &&& $\rightarrow$ TaSK &for Elderly Residents& \textbf{Scenic}: Two backdrops, Figurine, LEGO \\ &&&&& \textbf{Craft}: Pipe Cleaner  \\\midrule
    P4 & Guide & Parallel  &  Paper Sketching & Indoor Navigation & \textbf{Stage}: Spotlights\\ 
        &&& $\rightarrow$ TaSK &Companion for Elderly& \textbf{Scenic}: LEGOs, Blocks, Robot figurine  \\
        &&&&& \textbf{Craft}: Popsicles, Pipe Cleaner, Cards, Fur balls, Card, Sticky note \\ \midrule
    P5   & Companion & Parallel  &  Paper Sketching & Social play \& positive reinforcement to motivate & \textbf{Stage}: Bubbles, Fan, Turntable, Spotlight, LED strips, Sound Effects \\
        &&&$\rightarrow$ TaSK& children & \textbf{Scenic}: Blocks, Figurines \\
        &&&&& \textbf{Craft}: Felts, Popsicle \\ \midrule
    P6 & Companion &  Parallel & Paper Sketching & Competitive \& & \textbf{Stage}: Spotlight    \\ 
        &&& $\rightarrow$ TaSK &collaborative games to & \textbf{Scenic}: Blocks \\ 
        &&&&motivate children & \textbf{Craft}: Popsicles, Cards, Color papers  \\ \midrule
    P7  & Education & Sequential & Paper Sketching & Socially assistive robot & Did not use TaSK \\
        & $\rightarrow$ Guide&&&companion for elderly parent &\\ \midrule
    P8 & Education & Sequential & Paper Sketching & Office robot for & Did not use TaSK \\  & $\rightarrow$ Delivery&&&garbage disposal&\\ \bottomrule
    \Description{The result of the design session is shown in the table with six rows and six columns}
    \end{tabular}
\end{table*}

\subsection{Insight \#2: The Design Strategies}\label{sec:results:strategies}  

    In this section, we report the design strategies and approaches designers used to craft their HRI scenes with our proposed method (summarized in Table \ref{table:components_used}).     
    Notably, we observed that \textit{participant's diverse background and abilities} played a role in shaping their strategies and design process. For example, P1 drew from familiarity with using craft supplies in daily design practices. P3, drawing from expertise in kinetic architecture, incorporated dynamic elements into the scenario, such as a dynamic walker bar, highlighting the potential for design innovation within scene-based design. P4, who had prior experience visiting eldercare facilities, leveraged personal experiences to create a detailed and emotionally resonant eldercare scenario, or video game artist P6 designed scenarios with various game mechanics. 
    We present several common design strategies that were observed across the design session: 1) brainstorming the scene, 2) creating a scene flow, and 3) communicating the scenes. 
    \subsubsection{Strategies for Brainstorming the Scene: Paper Sketching, Note-taking and Tangible Design}

        There were four approaches observed for brainstorming a scene that included the use of TaSK or paper sketching. 
        P1 immediately was drawn to TaSK and \textit{only used TaSK} for creating their scenes. 
        P2 initially took time to explore each TaSK component and later supplemented their scene design with paper sketching. 
        Four participants (P3--6) sketched or took notes on paper first, then used the stage effect components, scenic components, and craft supplies from the kit to create their scenes.
        Among these four participants, P3 and P5 used the paper only to keep notes, while P4 and P6 used it to sketch the scenic layout from different perspectives. P4 sketched a top-down view and a first-person view and created these scenes with TaSK components. P6 also sketched a top-down view of their scene and later demonstrated it using TaSK components. 
        P7 and P8, however, only used paper-based sketching and note-taking approaches and did not use TaSK. 

    \subsubsection{Strategies for Creating a Scene Flow: Sequential and Parallel}

        Participants either created \textit{sequential} scenes or branched out and created \textit{parallel} scenes. Participants who created sequential scenes (P1-3-7-8) typically focused on one storyline, building a scenario from start to finish without introducing alternative scenes. Participants that branched out (P2-4-5-6) typically started with their chosen context, but later created several alternative scenes that branched out from the initial plot.
        
\begin{figure*}
    \centering
    \includegraphics[width=\linewidth]{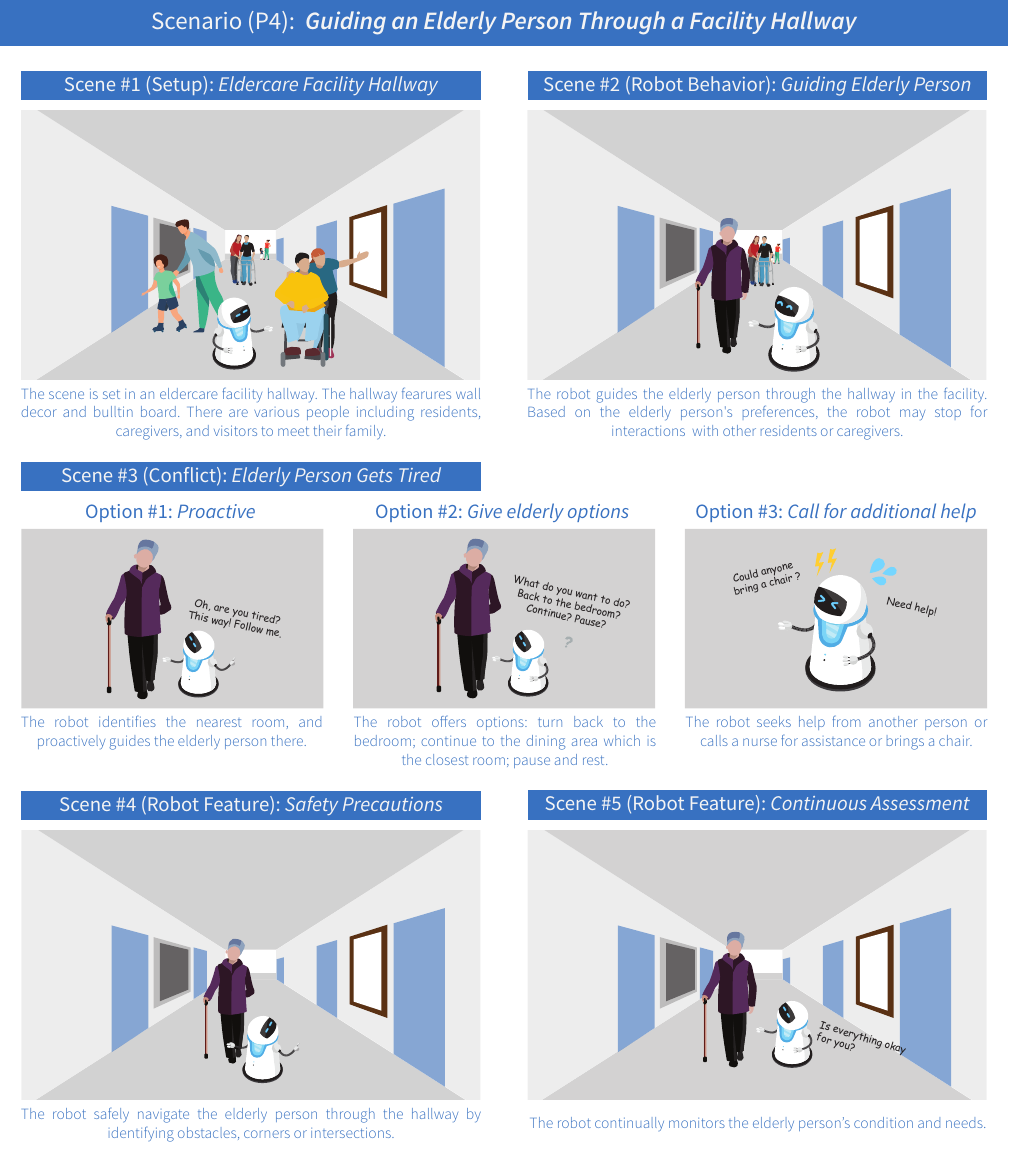}
    \caption{\textit{Scenario created by P4:} This figure shows a representation of P4's design process, supported by tangible scenography. Here we illustrate how scenes interact to create a holistic HRI scenario. P4 created multiple scenes with visual elements, robot behaviors (speech, movement, motivation, gestures), and parallel scenes that arose from the conflict. \textit{Copyright information:} Images by \texttt{pch.vector} and \texttt{macrovector} on Freepik}
    \Description{Seven illustrations in this figure. Each illustration shows a scene where a robot is working in a hallway in an eldercare facility. The illustration on the top left shows a robot in the center. Around the robot, there is a family visiting someone, a woman in a wheelchair, an old person with a walker, and two caregivers. The illustration on the top right shows a robot and an elderly person in the center. The illustrations in the middle show a robot and/or an elderly person. On the left, a robot is sweating and asking the person to follow a robot. In the illustration in the center, a robot raises its left arm and asks questions to the person. The right-most illustration shows a robot, raising both arms and showing signals to ask for help.
    The two illustrations on the bottom include a robot and an elderly person walking the hallway. The left illustration shows a robot moving ahead of the person. The illustration on the right shows a robot tuning into a person to ask a question.
}
    \label{fig:scenes-P4}
\end{figure*}

        We identified a variety of strategies between the participants that branched out. For example, three participants (P2-4-5) branched out \textit{after} the conflict in the plot was introduced, while P6 created multiple alternative scenes starting from their initial scene. 
        The purpose of these parallel scenarios also differed between each participant. P2 created four different arcs in their scenario for the \textit{robot's expressiveness} (active, passive, verbal and non-verbal expressiveness), P4 created three scenarios with different levels of authority for the \textit{robot's response style} (proactive, responsive, submissive)~(see Fig~\ref{fig:scenes-P4} illustrating P4's strategies), and P5 created two \textit{personas for the robot's} intervention (positive reinforcement and aversive conditioning). P6, however, created six parallel scenarios that focus on different \textit{game options}, two parallel scenarios that focus on \textit{strategies} the robot may employ towards children with different personas (\textit{personalization}), as well as \textit{modifications} of these scenarios based on different age groups or group sizes. After the conflict was introduced, P6 offered three strategies for the \textit{robot's behavior} (expressing emotion, diversion, and seeking parental assistance). Among all participants, P6 created the most amount of parallel scenarios within one scene-design session.

    \subsubsection{Strategies for Narrating the Scene: Tangible and Verbal}
\begin{figure*}
    \centering
    \includegraphics[width=\linewidth]{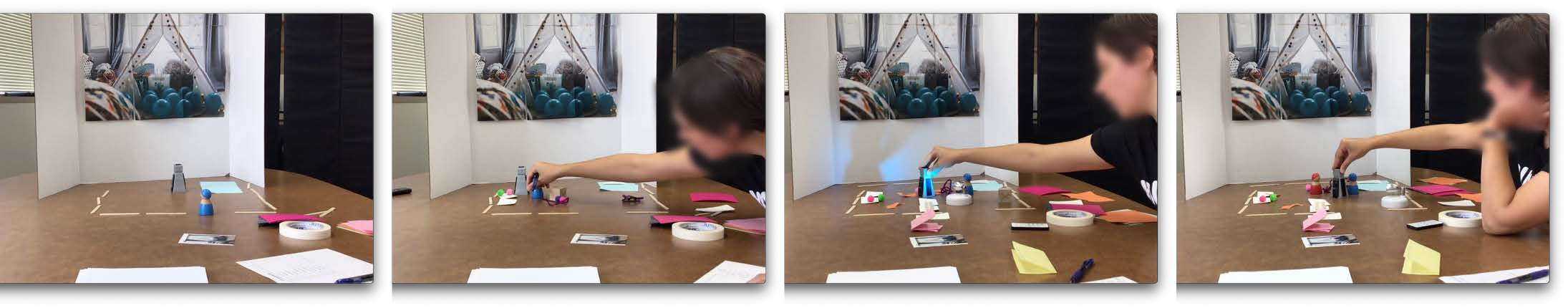}
    \caption{\textit{The creative process of tangible scenography.} P6 created a layout of a bedroom, placed robot-like and human-like figurines, created new props, and used lighting (blue). In this scenario, P6 demonstrates how a robot may help motivate children to clean up their room in a playful way.}
    \Description{Four photos are aligned. All photos include a backdrop on a table. On the front of a backdrop, there are some tangible components from TaSK. The first photo from the left shows two figurines and popsicle sticks on the table. The second photo from the left shows a person holding a figurine. The third photo from the left shows the person holding a blue-colored spotlight. The last photo shows the person holding a figurine.}
    \label{fig:process-P6}
\end{figure*}

        When describing their scene, TaSK components and our proposed method enabled \textit{tangible and active} strategies for narrating complex components of the scene, as an alternative to \textit{verbal and passive} strategies for narration. Specifically, we observed that four participants (P1-2-5-6) explicitly and continuously used the tangible components of the kit when describing their scenes (\textit{e.g.,} see Figure \ref{fig:process-P6}). Two participants (P3 and P4) narrated their scene with a mixed strategy, where they mostly shared verbal summaries, but occasionally used tangible resources to demonstrate their design. P7 and P8, however, did not use the tangible aspect of TaSK, and only verbally described elements of their scenarios.

\subsection{Insight \#3: Perceptions toward Tangible Scenography}

Interviews with participants revealed their preferences, challenges, and suggestions for applying tangible scenography as a method and using TaSK within their process.

\subsubsection{The Backdrop} \label{sec:results:backdrop} 
Three participants (P1, P2, P5) explicitly noted that the backdrop supported their process for setting the scene.
For instance, P1 highlighted the significance of creating scenes that resonate with the background, 
\textit{``the background helps. So you know if you're going to have, you know, a Walmart parking lot, would have been a different story.''} Participants appreciated how the backdrop influenced their conceptualization of the environment, with specific references to its ability to evoke certain narratives and scenarios. P5 described this as \textit{``definitely there's a lot of clutter in the scene. So this needs to be picked up (referring to objects in the backdrop). So this [backdrop] is setting the scene of clutter.''}

\subsubsection{Conflict Scenes} Secondly, participants were enthusiastic about the introduction of conflict within the design method which sparked creative tangents and drove the narrative forward. For example, four participants (P2, P4, P5, P6) branched out after the conflict and created parallel scenes (\textit{e.g.,} see Figures \ref{fig:scenes-P4} and \ref{fig:process-P6}). The presence of conflict not only inspired new scenarios but also encouraged designers to explore diverse perspectives and potential outcomes, thereby enriching the depth of their designs. 
P3 expressed that conflict was a familiar exercise within their current design practices, however, conflict as a part of tangible scenography proposed the \textit{added value of ``fun.''}

\subsubsection{Stage Effect Components} Lastly, the stage effect components in TaSK, such as fluid generation machines, spotlights, LED strips, a turntable, and sound effects, were instrumental in establishing the mood and atmosphere within scenes. 
While some participants utilized these components in alignment with their intended purpose (\textit{i.e.,} lighting and fog to express a mood), others explored alternative creative applications, leveraging them as props or functional elements within their designs. For example, P1 combined fluid generation machine (water) and turntable to create a fountain and explained their inspiration by saying \textit{``Honestly your contractions, they're awesome, but to me, I wasn't looking at the battery and all the wires and everything, I was really seeing a fountain.''}
  P4 described their desire to use stage effect components, but struggling to do so, which led them to diversify their design: \textit{``I wanted to use the bubble machine. But I couldn't figure it out. It felt out of place when I was imagining what I helped that eldercare facility would look like. So I thought about like, what is commonplace in other facilities? So I like went with that yellow-orange light, because that's what I like to associate them with.''}

\subsubsection{Challenges and Limitations of TaSK} However, challenges for incorporating TaSK were also evident, particularly for participants with traditional design backgrounds (\textit{e.g.,} P7 and P8). P8 expressed difficulty in adapting to the tangible format and preferred conventional sketching techniques:
\textit{``The idea of using materials like this is very foreign to me, especially (...) Yeah, I think I'm used to sketching my ideas and like using paper to explain things. Because I don't have very good spatial awareness as well.''}

\subsection{Insight \#4: Reflection on Facilitation Strategies} \label{sec:result:facilitation}

    We found that naturally emerging facilitation strategies, introduced as interventions in the sessions, were useful in guiding the design process.
    We observed four types of facilitator involvement enabled by tangible scenography: \textit{(1) summarizing the scene, (2) zoom in \& out, (3) spotlighting, and (4) modifying TaSK components} to support the participant's needs and preferences. 

    \subsubsection{Facilitation Strategy 1: Summarizing the scene}
    Given that scenes were built with abstract design components (\textit{e.g.,} spotlights, block figurines, craft supplies), it was important for facilitators and designers to have a shared understanding of the components as well as their purpose and role in the scene. To build this shared understanding, after the participant completed narrating their scene, the facilitator would point to each component, recite its role and purpose, and confirm with the participant. This strategy was used with all participants and enabled participants to confirm or elaborate on the components of the scene.

    \subsubsection{Facilitation Strategy 2: Zoom in \& out}
    Given that the scenes represented interconnected components and dynamic events, some participants were observed to either have too narrow or too broad a focus on the events in the scene. To balance this, the facilitator prompted the participant to adjust their focus by pointing out other components in the scene, asking curiosity-driven questions such as: \textit{``What is happening over here? Why is that specific color/sound/expression/figurine used? What happens next? How would the robot/human behave in response to that?''} This prompting style was used with all participants and allowed the participant to adjust their perspective by \textit{zooming in \& out} to describe the scene.

    \subsubsection{Facilitation Strategy 3: Spotlighting}
     After participants completed their design for human-robot interaction scenarios, they recorded a \textit{video} narrating their final design. In this process, facilitators applied \textit{spotlighting as a method} to support the storytelling of the final scenario. In this spontaneous method, the facilitator held the spotlight on the main actor being narrated by the participant to promote focused discussions. 

    \subsubsection{Facilitation Strategy 4: Customizing TaSK Components} 
    Facilitators often proposed spontaneous interventions to customize the use of TaSK components, as a response to the emerging needs expressed by the participants. For example, P3 expressed that they imagined a scenario taking place in two locations, an eldercare facility that opens up to an outdoor park. In response, the facilitators spontaneously offered to add both backdrops as part of their scene design. This enabled P3 to quickly demonstrate their envisioned scenario and encouraged them to explore how the robot's behavior would adapt in these dual settings. Furthermore, P7 and P8 expressed difficulty applying TaSK components within their scenario, which limited their ability to elaborate on the scenario holistically. Responding to this challenge, facilitators recommended spontaneous interventions to (1) change the selected scenario, (2) change the background, or (3) bring components from the kit to the scene. However, regardless of these interventions, participants were not able to create a complete scenario.

\section{Discussion}\label{sec:discussion}


\modified{Our work aimed to address three design challenges. \textbf{DC1:} What \textit{design approaches} can support creating holistic HRI scenarios that represent dynamic human environments and a wide range of robot interactions?
\textbf{DC2:} What \textit{design resources} may be useful in supporting the practices of creating holistic HRI scenarios?
\textbf{DC3:} How can these methods and resources facilitate the \textit{design process} of creating holistic HRI scenarios?}
To explore DC1, we adapted the concept of \textit{scenes}---individual environments characterized by their configurations of people, objects, spatial arrangements, and social norms---designed to demonstrate how robots may interact with people and their surroundings. 
To address DC2, we found inspiration from theatre production and scenography to develop the \textit{Tangible Scenography Kit (TaSK)} as a material resource to support scene-based design. The kit contained scenic components (backdrops, figurines, and blocks), stage effects (machines to generate fog, bubbles, and other water effects, spotlights, a turntable, and sound effects), craft supplies, and a design brief.
Finally, to explore DC3, we conducted exploratory design sessions with eight professional designers. The designers used \textit{TaSK} to create tangible and visual representations of their scenes. Our analysis suggests that tangible scenography holds the potential to help designers with little experience with robotic technologies in design ideation across a diverse set of interaction scenarios.
%
In the rest of this section, we reflect on takeaways from the scene-based design sessions and address the limitations and future directions.

\subsection{Scene-Based Design Can Support a Holistic Design Process} \label{sec:discussion:HolisticDesign}
\modified{Tangible scenography, as a scene-based design method, can support the creation of holistic human-robot interaction scenarios. We argue that, compared to traditional design approaches such as paper prototyping and storyboarding, tangible scenography can broaden the scope of design ideation and offer greater flexibility in narrative approaches. \modified{As noted in \S\ref{sec:results:process},} we observed that participants employed various techniques to initiate, develop, and convey their scenarios. Some participants immediately gravitated toward TaSK components, while others used a mix of strategies for sketching and brainstorming. 
Overall, tangible scenography involves manipulating a physical space to brainstorm and demonstrate how people and robots might interact in real-world settings. In a holistic design approach, this means considering how spatial arrangements and the changes in the environment may affect human and robot behavior and interaction flows. This spatial design perspective enables designers to consider interactions that go beyond the robot's task at hand. 
Tangible scenography also enables designers to weave several narratives into these physical spaces. Through these techniques, some participants were able to create environments that captured multiple narratives running in parallel. In other cases, participants focused on a sequence of events following a single narrative. By focusing beyond the robot and considering elements from the environment designers can brainstorm scenarios that better reflect dynamic nature of real-world settings.}

\subsection{TaSK Can Support Creative and Engaging Tangible Scenography} \label{sec:discussion:TaSK}
    The components of TaSK, including backdrops, stage effect components, scenic components, and the introduction of a conflict played a significant role in shaping participants' narratives. Participants highlighted the importance of these elements in both stimulating creativity and guiding their design process.

\subsubsection{Backdrops and Plots as Narrative Anchors}
Backdrops served as powerful narrative anchors that set the stage and provide context for scenes. Designers should consider the backdrop carefully, as it influences the direction and tone of the narrative. \modified{Our interview revealed that participants recognized its role in setting the scene, providing context, and influencing the direction of the narrative (reported in \S \ref{sec:results:backdrop}).} Backdrops not only served as visual cues but also supplemented storytelling and creative thinking. Backdrops created a space for designers to immerse themselves in a context, encouraging them to consider the nuances of the environment and its impact on human-robot interactions. 
Another pivotal component of scene-based design is the value of contextual plots that are key to setting the stage for interactions. Our study included several plots, such as patrolling in a park, guiding in an eldercare facility, and encouraging children to tidy their rooms. \modified{Each of these plots helped create a structure to guide and support designers' creative processes, as reported in \S \ref{sec:results:process}.}

\subsubsection{Stage Effects as a Resource for Dynamic Scenes}    
    In addition to the narrative anchors, the stage effects and scenic components in TaSK enabled participants to craft interactive and dynamic scenes. The \textit{stage effects}, such as spotlights, fog machine, and fan, allowed designers to manipulate lighting, atmospheric conditions, and sensory cues, adding depth and realism to their narratives. On the other hand, the \textit{scenic components}, which included figurines, craft supplies, and other various props provided a rich palette for designers to shape characters, objects, and context-specific details within each scene. \modified{Participants used the kit components in diverse ways to combine materials into new metaphors to support their narrative. For example, P1 prototyped a rotating water fountain combining the turntable and water machine; P2 represented pollutants at the park and a rotating base for the robot; P4 created bulletin boards and art decor; and P5 and P6 created various play contraptions.
    Furthermore, \textit{tangible narration} added an interactive dimension to the design sessions, similar to the robot programming environments presented by \citet{porfirio2021figaro}, allowing designers to physically manipulate components in the scene, visualize spatial relationships, and tangibly demonstrate robot behaviors. Together, these components facilitated dynamic representations of robot behaviors, user interactions, and environmental changes, allowing participants to convert abstract concepts into tangible and visually compelling HRI scenarios. }

\subsubsection{Conflict as Creative Catalyst to Promote Engagement} Introducing a new conflict in the scenario acted as a catalyst for creativity. Conflicts enabled designers to explore alternative robot behaviors and strategies, leading to more dynamic and multifaceted narratives.
\modified{Consequently, we identified conflict as a key factor for design. The introduction of a conflict within the scenario also provided a milestone in the design process, which might be a useful strategy to introduce more structure to research-through-design practices~\cite{luria2019championing}. Conflict elicited creative new ideas and enabled participants' ability to branch scenarios out to diverse directions, \modified{as presented in \S \ref{sec:results:process}}. Participants responded to the presented conflict as an opportunity to explore alternative robot behaviors and strategies, resulting in more dynamic and multifaceted narratives. Furthermore, the conflict introduced a fun and playful new element to the process, making it engaging for designers. We observed that tangible scenography has the potential to create an avenue for designers to harness \textit{playfulness}, which can facilitate curiosity-driven, imaginative, and enjoyable design sessions.}
    
    \subsubsection{Designer Backgrounds Shaped Scene-Based Design Practices}
    We suggest that the tangible components in TaSK, as opposed to paper-based approaches, can lower the barrier to entry into HRI design. TaSK can support a more intuitive design approach for designers who are novices in the HRI field. For curating TaSK components (captured in Section \S\ref{sec:designprocess}), we placed a specific focus on identifying effective, simple, yet versatile resources to support scene-based design. Our goal was to provide a more immersive and engaging way for designers to convey complex interactions and dynamic changes in the environment. Eventually, we observed that the TaSK components encouraged designers to create an assortment of physical scenes. 
\modified{
    However, we did not anticipate the extent to which the combination of designers' backgrounds and TaSK components would foster the creative design process and the emergence of imaginative design ideas.
    In the design sessions, participants crafted novel use cases that combined TaSK components to supplement their scenes. Each designer brought their unique background and expertise to contribute to the richness and depth of their creative exploration. Through this process, participants leveraged the tangible elements of TaSK to prototype, iterate, and refine their ideas in real time, allowing for immediate reflection and adjustments. 
    For example, P3, drawing from expertise in kinetic architecture, incorporated dynamic elements into the scenario, such as a dynamic walker bar, highlighting the potential for design innovation within scene-based design. P4, who had prior experience visiting eldercare facilities, leveraged personal experiences to create a detailed and emotionally resonant eldercare scenario. Furthermore, video game artist P6 designed scenarios with various game mechanics. As designers drew from their diverse professional and personal experiences in fields such as landscape architecture design~(P3), book editing~(P5), and video game design~(P6), they infused their ideas with a variety of influences and insights, resulting in creative solutions. 
     }
     \subsubsection{Facilitation Strategies Matter}
    The influence of designers' backgrounds and their approach to utilizing TaSK resources facilitated a dynamic exchange of perspectives and approaches with facilitators, fostering an environment of innovation and exploration. The facilitators were able to support the designers' iterative process by offering strategic solutions to utilize the TaSK components. In our exploratory study, we observed that facilitation strategies played an important role in supporting the participants' creative processes. \modified{As reported in \S\ref{sec:result:facilitation}, \textit{summarizing} scenes helped build a shared understanding of the scenarios that designers created. \textit{Zooming in and out} encouraged participants to consider various perspectives and components within the scenes, ensuring a holistic exploration of the scenario. Other interventions such as \textit{spotlighting} or allowing for \textit{customization of TaSK components} addressed emerging needs expressed by the participants. Hence, we emphasize added value of facilitators in guiding the scene-based design process. Facilitators should be prepared to provide support by summarizing scenes, prompting designers to explore different perspectives, and making spontaneous interventions when needed. Furthermore, these facilitation strategies may serve as a semi-structured design technique, addressing challenges noted by~\citet{luria2019championing} regarding the subjective and unstructured nature of RtD methods.}


\subsection{Takeaways for Practitioners}

\modified{
    Tangible scenography can offer several benefits to HRI practitioners. In Section \S\ref{sec:discussion:HolisticDesign} and Section \S\ref{sec:discussion:TaSK}, we motivated and discussed \textit{why} tangible scenography can be of value to practitioners.}
\modified{    
    In this section, we discuss three examples for \textit{when} and \textit{how} tangible scenography can be applied in the HRI context, across different phases of design. 
    To discuss the examples, we will introduce and refer to the framework for \textit{Universal Methods for Design (UMD)} \cite{hanington2019universal}. UMD proposes five phases of design, however, we will specifically focus on the first three stages: Phase 1 --- Planning, Scoping, and Definition; Phase 2 --- Exploration, Synthesis, and Design Implications; and Phase 3 --- Concept Generation and Early Prototype Generation. Because we position tangible scenography as an early-phase design and exploratory method for brainstorming, we will not discuss use cases for Phase 4 --- Evaluation, Refinement, and Production, and Phase 5 --- Launch and Monitor. However, future work may expand and explore applications of tangible scenography tailored for those later design stages.}
\modified{    
    \subsubsection{Phase 1 --Planning, Scoping, and Definition}
    This UMD phase involves the exploration of design parameters, clarifying the scope of what needs to be designed. Designers can utilize TaSK components, particularly scenic components, to represent contextual factors present in the scene, such as stakeholders or environmental parameters. Through the process of setting the scene, designers can determine which design factors to prioritize.
    For example, setting a scene for \textit{``a robot patrolling in a park''} reveals contextual factors like ``when'' or ``who.'' The population in the park may vary depending on the time of day, prompting designers to consider ``time'' as a key design factor. Additionally, as the park is outdoors, the designer can brainstorm sequential scenes where weather conditions might change. These contextual factors may also act as an important parameter for several robot design requirements, including \textit{material}, e.g., ``what are the design requirements for waterproofing the robot's exterior;'' \textit{hardware}, e.g., ``what sensors are necessary to detect weather changes;'' or \textit{interaction}, e.g., ``what is the protocol to follow when there is a bad weather alert?'' By identifying such design requirements, designers and engineers can collaboratively progress through the decision-making process to determine which factors should be the focus of their attention.}
\modified{
    \subsubsection{Phase 2 --Exploration, Synthesis, and Design Implications} This UMD phase explores design implications by \textit{``immersive research and design ethnography~ \cite{hanington2019universal}.''} Tangible scenography enables designers to create a detailed scene, which immerses them in the context. By physically arranging these elements, designers can explore design implications in the scene.
    For instance, by setting a scene, building a scene after introducing a conflict, or even narrating scenes, designers may be able to realize narrow spaces, crowded areas, or blind spots, which prompts design considerations for robot navigation (\textit{i.e.,} the robot should be able to recognize and adjust the speed to ensure safety) and communication (\textit{i.e.,} the robot should communicate with people nearby to make sure that its path is clear).}
\modified{   
    \subsubsection{Phase 3 --Concept Generation and Early Prototype Generation} This UMD phase aims to generate a concept and create an early prototype through participatory and generative design activities. TaSK components can be used not only for setting the scene but also materializing the ideas. In addition, tangible scenography allows designers to collaborate in a shared environment, where they can collaboratively brainstorm and iteratively experiment with their prototype in real-time.
    For example, a designer may be involved in co-designing a robot to guide patients in an eldercare facility. The tangible scenography session can accommodate multiple participants, such as a designer, a facilitator, an elderly patient, and their caregiver. During the session, the designer's goal is to propose an idea to help elderly patients navigate from an indoor location to an outdoor facility. The designer might use the kit components such as a backdrop to set the scene and robot figurines to demonstrate their design ideas. At this point, the caregiver may intervene and point out to safety risks and design the layout of the care facility using craft supplies. This, in turn, can promote a shared understanding between the designer, caregiver, and elderly patient. The elderly patient might intervene and propose a new conflict depending on their personal experiences and preferences. These spontaneous, cooperative interactions among co-designers can contribute to the evolution of an early prototype and concept generation. Overall, by applying tangible scenography as a co-design method, participants can utilize TaSK components and iteratively experiment with different design ideas.}


\modified{
    In summary, these three examples illustrate the diverse applications and benefits of tangible scenography across three different UMD phases. By providing a semi-structured and immersive design environment, tangible scenography empowers HRI practitioners by helping them communicate their ideas effectively, engage in collaborative brainstorming, and navigate decision-making to develop robot design requirements.}

    \subsection{Limitations and Future Directions }
    Our work has a number of limitations. 
    First, our process for creating TaSK is limited by a single expert's view. \modified{We only interviewed one expert due to our limited access to local experts and practical limitations for in-person interviews. We acknowledge that the single expert's process may not represent the general practices of the field of theatre and scenography and may constrain the breadth of our supplemental design components. }
    However, this process contributed as an inspirational starting point to iteratively create a design kit. We will provide TaSK components as open source to enable its replication and refinement.
    Second, TaSK components may be limited in accommodating \textit{diverse needs of designers}. We saw that participants operationalized tangible scenography and TaSK components in various ways. Some participants relied more on verbal descriptions and paper-based sketches. However, two participants (P7 and P8) found it difficult to operationalize the TaSK components. \modified{This observation suggests that the method and TaSK components are limited in generalizing to various designers' needs.} 
    Future iterations of TaSK could benefit from enhancements that make it more accessible and intuitive for designers coming from varying backgrounds. A possible solution could include encouraging participants to bring their familiar design materials to the sessions which might help participants to be unblocked. 
    Furthermore, future versions of TaSK can combine digital and tangible resources, including a projected backdrop to represent dynamic changes in the scene. TaSK can include clothing and costumes for the robot figurines (similar to \cite{friedman2021designing}). 
    Third, as discussed above, practitioners of tangible scenography may have various needs and skills. This method might extend to serve the needs of specific design partners, such as special populations with diverse accessibility needs (\textit{e.g.,} children or adults with cognitive disabilities, or older adults). This method may provide a tangible and intuitive way to communicate needs and co-design robot behaviors and movements as an alternative to other methods such as storyboarding, sketching, and using post-it notes for brainstorming. 
    Moreover, having a \textit{single session} of tangible scenography in this study may be limiting, such that sessions focused on high-level discussions of scenes rather than the in-depth design of robot behaviors.

\section{Conclusion}
Drawing inspiration from the performing arts, we propose \textit{tangible scenography}, a holistic and exploratory design method to craft scenarios that capture human-robot interaction's complex and dynamic nature, and \textit{Tangible Scenography Kit (TaSK)} as a supportive artifact for this method. It encourages designers to think beyond the functionality and behaviors of a robot and consider the social aspects of robot interactions, as well as the surrounding factors that may emerge in these complex and dynamic environments.
While our study provides an initial exploration of this method, further research and refinement are still necessary to extend its use across domains and populations. As the field of design practices in HRI continues to evolve, we argue that \textit{tangible scenography} may stand as a creative and valuable addition to the designer's toolkit.

\begin{acks}
We would like to acknowledge Batuhan Bayraktar for his efforts in contributing to TaSK and Pragathi Praveena for her valuable discussions in synthesizing our findings.
\end{acks}

\balance
\bibliographystyle{ACM-Reference-Format}
\bibliography{references}

\end{document}